\documentclass[runningheads]{llncs}
\usepackage{graphicx}
\usepackage{booktabs, multirow}
\usepackage[table,xcdraw]{xcolor}
\usepackage{caption}
\usepackage{subfigure}

\begin{document}
\title{Cascaded Volumetric Convolutional Network \\ for Kidney Tumor Segmentation \\ from CT volumes}
%
%
\author{Yao Zhang\inst{1, 2} \and Yixin Wang\inst{1, 2} \and Feng Hou\inst{1, 2} \and Jiawei Yang\inst{3} \and Guangwei Xiong\inst{3} \and Jiang Tian\inst{4} \and Cheng Zhong\inst{4}}
\authorrunning{Zhang et al.}
%
\institute{Institute of Computing Technology, Chinese Academy of Sciences \and
University of Chinese Academy of Sciences \and
Southeast University \and
AI Lab, Lenovo Research\\
\email{zhangyao215@mails.ucas.ac.cn, wangyx57@lenovo.com, houfeng1@lenovo.com, xionggw1@lenovo.com, yangjw3@lenovo.com, tianjiang1@lenovo.com, zhongcheng3@lenovo.com}}
%
\maketitle              
\begin{abstract}
Automated segmentation of kidney and tumor from 3D CT scans is necessary for the diagnosis, monitoring, and treatment planning of the disease. In this paper, we describe a two-stage framework for kidney and tumor segmentation based on 3D fully convolutional network (FCN). The first stage preliminarily locate the kidney and cut off the irrelevant background to reduce class imbalance and computation cost. Then the second stage precisely segment the kidney and tumor on the cropped patch. The proposed method ranks the $4th$ place out of $105$ competitive teams in MICCAI 2019 KiTS Challenge with a Composite Dice of 90.24\%.

\end{abstract}
\section{Introduction}
There are more than 400,000 new cases of kidney cancer each year, and surgery is its most common treatment. Due to the wide variety in kidney and kidney tumor morphology, there is currently great interest in how tumor morphology relates to surgical outcomes, as well as in developing advanced surgical planning techniques. Automatic semantic segmentation is a promising tool for these efforts, but morphological heterogeneity makes it a difficult problem. 


To tackle these difficulties, many segmentation methods have been proposed. Recently, 2D deep neural networks based on Fully Convolutional Network (FCN) have been applied with success to natural images and also in medical imaging~\cite{ronneberger2015u}. However, 2D FCN based methods ignore the contexts on the z-axis, which would lead to limited segmentation accuracy. To solve this problem, many researchers proposed to use 3D deep neural networks to probe the 3D contexts~\cite{cicek20163d}~\cite{milletari2016v}. And experimental results showed that 3D FCNs generally achieve better performance than the 2D FCNs in different organ segmentation tasks. But compared to 2D FCNs, 3D FCNs suffer from high computational cost and GPU memory consumption. The high memory consumption limits the depth of the network as well as the receptive field, which are the two key factors for performance gains. So it is difficult for researchers poor in GPU memory to improve the performance of tumor semantic segmentation. Meanwhile, the kidney only occupies a small region of the abdomen and the kidney tumor is even more smaller. Such an imbalance of kidney, tumor and background results in the difficulty in training a segmentation model.





In this paper, we propose a cascaded segmentation framework using volumetric fully convolutional networks for MICCAI 2019 Kidney Tumor Segmentation Challenge (KiTS 2019). It consists of two main stages: one for coarse kidney localization and the other for accurate kidney and tumor segmentation. In the first stage, we use a segmentation based localization strategy to estimate a region that covers the whole kidney, and leave out pixels outside to alleviate class imbalance and cut down memory consumption; In the second stage, we train a fine segmentation network based on the cropped kidney region obtained in the first stage, and then transform the predicted masks in target region to the original size volume. In these two stages, volumetric fully convolutional neural networks adapted from 3D UNet are used to automatically segment the kidney and tumor. Moreover, the mask of kidney from the first stage is also fed into the second stage with CT scans as a spatial prior, which further improves the performance of the fine segmentation model.

\begin{figure}[!tp]
\includegraphics[width=\textwidth]{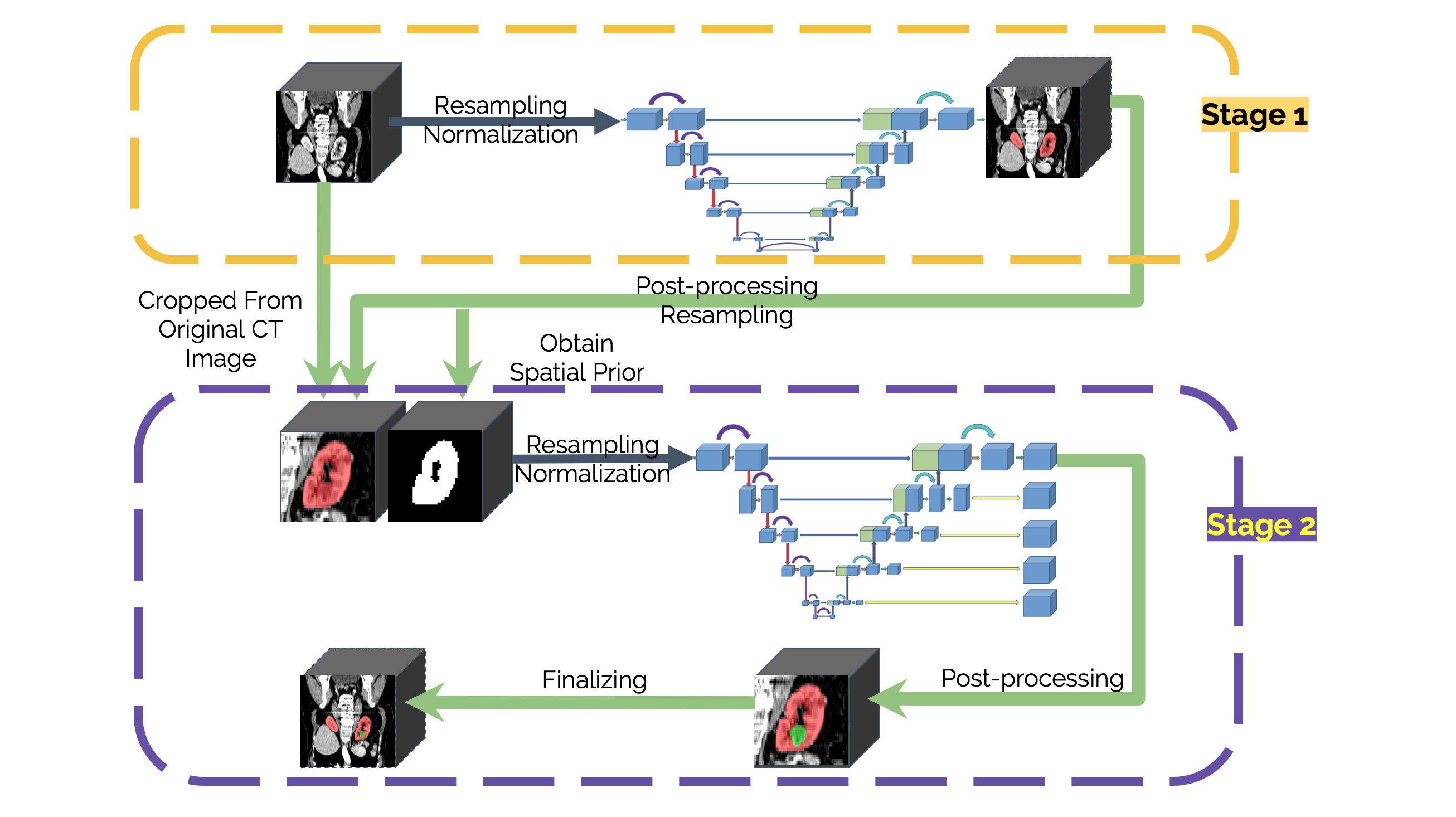}
\caption{The pipeline of kidney and tumor segmentation.} 
\label{pipeline}
\end{figure}

\section{Method}
\label{sec:method}
\begin{figure}[!tp]
\includegraphics[width=\textwidth]{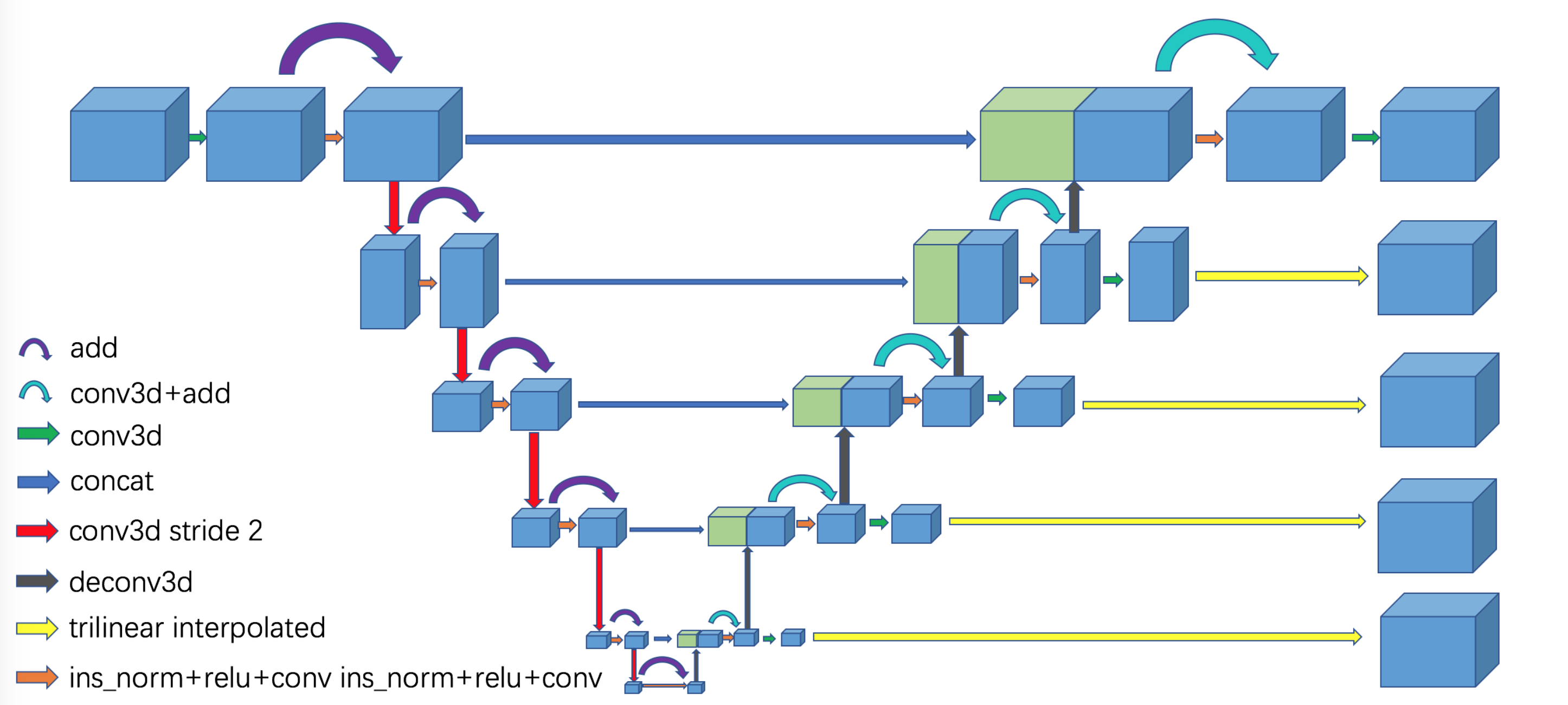}
\caption{The kidney and tumor segmentation network.} 
\label{network}
\end{figure}

Since the kidney takes only a very small fraction of the entire CT image, its segmentation is easily misled by irrelevant tissues. In addition, the tumor appears on kidney in various shape and position. Such class imbalance leads to extreme hard recognition and segmentation. Therefore, we design a two-stage method based on 3D UNet model, as illustrated in Fig.~\ref{pipeline}. The first is a basic model to localize the kidney and crop sub-volume patches cover the whole kidney. The second is a multitask 3D UNet model to segment kidney and tumor simultaneously based on patches from the first stage. The details of model design are shown in Fig.~\ref{network}.

\subsection{Kidney Localization}
In order to improve the classification performance and prediction ability on the imbalanced dataset of kidney and tumor, we adopt 3D U-Net model as our framework to first locate kidney. For 3D image data, we train such a model to aggregate valuable information along the z-axis.

This model is a 3D fully convolutional neural network based on UNet architecture. It takes 3D volumes as input and processes them with 3D operations. Due to the large size of kidney patient's image, the input patch size is set as $80\times160\times160$ and batch size as 2. The left part is an encoding path containing five layers. Each layer contains two $3\times3\times3$ convolutions each followed by instance normalization and leaky ReLUs activation instead of popular batch normalization and original ReLU in 3D UNet. Then a $1\times2\times2$ or $2\times2\times2$ max pooling with strides of two is achieved. At the layer with the highest resolution, 30 feature maps is used and doubles by the layers. Considering the patch size, we set the times of pooling operation per axis as (4, 5, 5) until each axis of the feature map is smaller than 8. The decoding path on the right is a symmetric counterparts of the left recovering spatial information. Each layer consists of convolution with size $3\times3\times3$, normalization and leaky ReLU operation to receive semantic information and skip connection to recombine it with essential high-resolution features from encoding path, which catches sufficient contextual information about kidney. The up-sampling is achieved by transposed convolution operation. After this stage to localize kidney preliminarily, we then crop a series of 3D volume patches and pass them to the second stage for further fine segmentation of kidney and tumor.

\subsection{Kidney and Tumor Segmentation}

Similar to kidney localization network, our model is composited of encoder part and decoder part and takes the advantages of other enhancements, which are connected via some residual connections. The input patch size of kidney and tumor segmentation is set as $40\times128\times128$ and batch size as 5. Most variants of UNet modify the original basic block in each layer and the skip connection construction manner. Here, from different inspirations, we aggregate the pre-activation and residual connection from ResNet~\cite{he2016deep} and deep supervision loss~\cite{Dou20163D} from 3D U-Net with Multi-level Deep Supervision with a fine-constructed U-Net work to form our model.

The encoder part uses pre-activation ResNet blocks, where each block consists of two convolutions with the instance normalization and ReLU as activation function. Skip connection is used after each blocks. We follow a common CNN approach to progressively downsize image dimensions by 2 and simultaneously increase feature size by 2. For downsizing we employ strided convolutions. All of  the convolutions are $3\times 3\times 3$ except the first convolution with the shape $1\times 3\times 3$  to balance the spacing . The number of pool operations in each axis are 3, 5, 5 based on the analysis of the dataset. The base number of filters is set to 30.

The decoder is similar to the encoder. At the beginning of each special level, we use deconvolution to double the spatial dimension and reduce the number of features, followed by an addition of encoder output through the skip connections. At the end, we apply a pre-activation ResNet blocks like the encoder. But we reduce the feature size by 2 in the first convolution and use a 1x1x1 convolutions before residual connection to make sure they have the same shape. Besides, we apply deep supervision to enhance the discriminative ability of medium-level features. In each decoder level, we use a convolution and an upsampling to get the same spatial size as the original image and calculate the loss using the masks from all level.

To further increase the segmentation performance and robustness, the final result is aggregated by ensembling.

\subsection{Pre-processing}
Due to the limited numbers of data cases, careful preprocessing is necessary for fully utilizing KiTS2019 dataset. Interpolated data with the same affine transformation over all patients from that dataset are used in this work. The motivation behind choosing preprocessing schemes and collecting selected statistics is to capture the basic distribution information over all the data in this dataset. This work mainly follows the nnUNet~\cite{Isensee2018nnU} pre-processing path but deprecates some trivial parts like cropping with regard to Kits2019 dataset.
\paragraph{Resampling}

In this work, resampling scheme firmly follows the idea of preserving the information from the low resolution axis as much as possible while completely exploiting GPUs memory. 

For stage 1, the extraction of region of interest (ROI), namely the kidney and kidney tumor area extraction, we downsample the 2D planes of the original images along the slice axis, which has the lowest resolution, by doubling the slice axis voxel spacing. That will compatibilize resolutions along three dimensions so that the anisotropy problem can be ameliorated. In addition, a patch size of $80 \times 160 \times 160$ that fits our GPUs can now have a receptive filed more than a quarter of total voxels in the resampled images. Having more contextual information over entire volume is believed to improve model learning when using patches. 

For stage 2, the segmentation of kidney and kidney tumor, no resampling is needed for its unnecessity. The output sizes of ROI extraction have been dramatically reduced compared to the original ones (median size of $40 \times 160 \times 160$ v.s. median size of $138 \times 512 \times 512$). Feeding entire cropped volume into the model can better utilize both GPUs memory and data. Along with the CT volume, the kidney segmentation is also concatenated as the input of stage 2. With the guidance from the kidney mask, the burden in searching for the optimal solution in the precise segmentation of kidney and kidney tumor has also been well alleviated.

\paragraph{Normalization}
Normalization is performed separately in each entity of the dataset, whose data vary greatly in the range of intensity scale and HU values. The inputs of both stages are normalized in the same manner: for each data case, 1) do percentile clipping of (0.05\%, 99.5\%) to remove outliers and 2) do standardization by subtracting the mean and standard deviation values collected in statistics collecting step. Normalization over entire image in stage 1 benefits ROI extraction and normalization over solely ROI enhances learning targets so as to facilitate model learning.

\subsection{Postprocessing}
Given by the prior knowledge that no more than two kidneys exists in the abdomen, the segmentation results are followed by a connected component analysis to remove false positives. In the first stage, the largest two components are retained; while in the second stage, the largest component is retained and the rest are discarded.

\section{Experiments}
\label{sec:experiments}


We evaluate our method on the competitive dataset of MICCAI 2019 KiTS Challenge. It is a collection of late arterial phase CT imaging, segmentation masks, and comprehensive clinical outcomes for 300 patients who underwent nephrectomy for kidney tumors between 2010 and 2018. The dataset contains 210 and 90 contrast-enhanced 3D abdominal CT scans for training and testing respectively. We select 50 CT scans from training set for validation and the remained 160 CT scans for training. We implemented our network in Pytorch and trained it on an NVIDIA Tesla V100 32GB GPU. We train the network from scratch. The Adam optimizer with an initial learning rate of 0.0003 is used for parameters updating. The loss function is a summation of Cross Entropy Loss and Dice Loss. It took about 18 hours for training the model in the first stage and about 30 hours for it in the second stage. The results on the validation set are listed in Table~\ref{kits}. We also evaluate our method on the online test set held out by the organizers. Table~\ref{kits_test} reports the results, which can also be found at http://results.kits-challenge.org/miccai2019/.

\begin{table}[!tp]
\renewcommand{\arraystretch}{1}
\centering
\caption{Quantitative results of kidney and tumor segmentation on validation set. First two raws are the results in the first stage and the last five raws are the results in the second stage. DS, Res and SP indicate Deep Supervision, Resblock and Spatial Prior respectively.}
\label{kits}
\setlength{\tabcolsep}{10mm}{
\begin{tabular}{@{}c|c|c@{}}
\toprule
\multicolumn{1}{c|}{\textbf{Methods}} & \textbf{Kidney Dice {[}\%{]}} & \textbf{Tumor Dice {[}\%{]}} \\ \midrule
\textbf{2D UNet}                  & 95.56   &       62.35           \\
\textbf{3D UNet}                   & 97.70   &        73.31           \\ 

\midrule
\textbf{2D UNet}                  & 96.37   &       67.75            \\
\textbf{3D UNet}                   & 97.24   &      78.23             \\
\textbf{3D+DS}                  & 97.75   &       80.29            \\
\textbf{3D+DS+Res}                  & 97.66   &     83.49              \\
\textbf{3D+DS+Res+SP}                  & 98.05   &    83.70               \\

\bottomrule
\end{tabular}}
\end{table}

\begin{table}[!tp]
\renewcommand{\arraystretch}{1}
\centering
\caption{Top 5 results of kidney and tumor segmentation on KiTS2019 online evaluation (http://results.kits-challenge.org/miccai2019). Teams were ranked by Composite Dice. }
\label{kits_test}
\setlength{\tabcolsep}{0.5mm}{
\begin{tabular}{@{}c|c|c|c|c@{}}
\toprule
\multicolumn{1}{c|}{\textbf{Team}} & \textbf{Kidney Dice} & \textbf{Tumor Dice} & \textbf{Composite Dice} & \textbf{Rank} \\ \midrule
\textbf{Fabian Isensee et al.}                  & 97.37   &       85.09     &     91.23    &     1      \\
\textbf{Xiaoshuai Hou et al.}                   & 96.74   &      84.54      &      90.64      & 2 \\
\textbf{Guangrui Mu et al.}                  & 97.29   &       83.21   &       90.25   & 3        \\
\textbf{Yao Zhang et al. (Ours)}                  & 97.42   &     83.06   &     90.24   & 4        \\
\textbf{Jun Ma}                  & 97.34   &    82.54    &    89.94  & 5         \\

\bottomrule
\end{tabular}}
\end{table}

\section{Conclusion}
\label{sec:conclusion}

In this paper, we design and develop a cascaded framework for automatic kidney and tumor segmentation from CT scans, which consists of two volumetric fully convolutional networks. The first network is used to coarsely segment the kidney from a low-resolution input and estimate the location of the kidney. The second is used to further precisely segment the kidney and tumor from the cropped patch that covers the whole kidney. Furthermore, we enhance the FCN with residual link, deep supervision and spatial prior.

%
%
%
%
%
\bibliographystyle{splncs04}
\bibliography{mybibliography}

\end{document}